\documentclass[aps,preprint,groupedaddress]{revtex4}
\usepackage{amssymb}
\usepackage{amsbsy}
\usepackage{rotating}
\usepackage{amsmath}
\usepackage{rotating}
\usepackage[latin1]{inputenc}
\usepackage{amsthm,mathrsfs,amsopn}

\usepackage{graphicx}

\begin{document}

\title{Linear noise approximation for stochastic oscillations of intracellular calcium}

\author{Laura Cantini$^{1}$, Claudia Cianci$^{2}$, Duccio Fanelli$^{3}$, Emma Massi$^{1}$, Luigi Barletti$^{1}$}
\affiliation{
1. Dipartimento di Matematica e Informatica ``U.\ Dini'', University of Florence, Viale Morgagni 67/A, 50139 Florence, Italy\\
2. Dipartimento di Sistemi e Informatica and INFN, University of Florence, Via S. Marta 3, 50139 Florence, Italy\\
3. Dipartimento di Fisica e Astronomia, University of Florence and INFN, Via Sansone 1, 50019 Sesto Fiorentino, Florence, Italy}

\begin{abstract}
A stochastic model of intracellular calcium oscillations is analytically studied. The governing master equation is expanded under the linear noise approximation and a closed  prediction for the power spectrum of fluctuations analytically derived.  
A peak in the obtained power spectrum profile signals the presence of stochastic, noise induced, oscillations which extend 
 also outside the region where a deterministic limit cycle is predicted to occur. 
\end{abstract}

\maketitle

\vspace{0.8cm}

\section{Introduction} 
\label{S1}

Calcium $Ca^{2+}$ oscillations prove fundamental in many different life processes, including, among the others,  muscle contraction,  neural activity and fertilization \cite{cell}. At rest conditions, the calcium in the cell cytoplasm is kept at low concentration, while it is present at much higher concentration outside the cell, or inside small intracellular compartments as the endoplasmic reticulum, the sarcoplasmic reticulum and the mithocondria. Large gradients can indeed induce a sudden increase in the concentration of calcium dispersed inside the cellular {\it milieau}, by either releasing it from the internal stores or importing it from the outside environment, through specific channels. 

In non-excitable cells, binding of an agonist, hormone or neurotransmitter, to cell-surface receptors initiates a cascade of reactions which promotes the production of the second messenger inositol trisphosphate (IP3). This latter diffuses through the cytoplasm and eventually binds to the IP3 receptors, positioned on the membrane of the endoplasmatic reticulum. The IP3 receptors act also as channels: upon binding of the IP3, the channels open and let the $Ca^{2+}$ to flow from the 
endoplasmatic reticulum into the cell cytoplasm. Importantly, the release of calcium as mediated by the IP3 receptors can occasionally stimulate an additional release of $Ca^{2+}$ from the endoplasmic reticulum. This is an autocatalytic process, usually termed calcium-induced calcium release (CICR)  \cite{review}. 

Different models have been developed in the past to describe the self-consistent generation of calcium oscillations. 
According to the pioneering model \cite{Meyer} sustained oscillations of
cytosolic $Ca^{2+}$ develop as mediated by the rise in IP3, triggered by
external stimulation. This rise elicits the release of $Ca^{2+}$ from an
IP3-sensitive intracellular store, a process which in turn activates a further release of calcium from a second, independent compartment insensitive to IP3.
Building on this formulation Goldbeter and collaborators \cite{goldbeter1,goldbeter2} have then elaborated a simplified scheme, particularly interesting for pedagogical reasons, where two distinct species, the cytosolic calcium and the calcium stored inside a IP3 sensitive compartments, are solely assumed to mutually interact. Working in this simplified setting, it was observed \cite{goldbeter3}  that repetitive calcium spikes, evoked by external stimuli, are not necessarily linked to concomitant IP3 oscillations. The models mentioned above are deterministic in nature and, as such, assume the system to be ideally described in terms of continuum concentration amounts. 

As opposed to this vision, one can favour an individual based description, which effectively accounts for the intrinsic
discreteness of the scrutinized system. Stochastic effects are therefore present and stem from the finite size of the population of elementary constituents. Such stochastic contributions, endogenous to the system, can amplify via a resonant mechanism and so yield macroscopic oscillations in the discrete concentration in a region of the parameters for which a stable fixed point is predicted, as follows the deterministic linear stability analysis \cite{McKanePRL,dipatti}. Similar conclusions apply to spatially extended systems \cite{McKane,De_Anna,Goldenfeld,Biancalani_2010}. As concerns calcium dynamics, stochasticity has been mainly associated to external disturbances  \cite{Marchant,Falcke}. The stochastic opening and closing of the channels have been for instance identified as a plausible cause of perturbation. At variance, Li, Hou and Xin \cite{Li-Hou} have recently investigated a discrete version of the Goldbeter 
model, showing that the inherent demographic noise can possibly drive stochastic oscillations, even when the underlying deterministic system is in a non-oscillatory state. The analysis carried out in \cite{Li-Hou} relies on numerical simulations. 
However, the effect of finite size fluctuations can be also analytically appreciated by expanding the governing master equation under the so called Linear Noise Approximation scheme. In doing so, one can obtain a close prediction for the power spectrum of stochastic fluctuations and characterize the resonant frequency as a function of the parameters of the model. 

In this paper we shall work along these lines, by revisiting the microscopic model \cite{Li-Hou}, which we will slightly modify. We will then perform a complete analytical treatment of the model, recover the Goldbeter's scheme in the mean field limit and characterize the distribution of stochastic fluctuations in terms of a Fokker-Planck equation derived from first principles. The stochastic oscillations observed in \cite{Li-Hou} are here interpreted as {\it quasi-cycles} of the discrete microscopic model, so building an ideal bridge with e.g. \cite{McKanePRL}. 

The paper is organized as follows. In the next section we will introduce the stochastic model, inspired to \cite{Li-Hou} and constructed so to converge to the Goldbeter scheme \cite{goldbeter1,goldbeter2} in the deterministic limit. 
In section \ref{S3} we will then turn to study the governing master equation, under the linear noise approximation and 
elaborate on the role of stochastic fluctuations. We will in particular obtain a close prediction for the power spectrum of fluctuations that we will benchmark to direct simulations. An approximate expression for the resonant frequency is also derived and proved to be adequate. Finally, in Section  \ref{S4} we will sum up and draw our conclusions.

\section{Stochastic model and the master equation} 
\label{S2}

We will hereafter introduce the stochastic a model for intracellular calcium oscillations that we shall study throughout the paper. The model describes the process of calcium-induced calcium release (CICR), a biological process whereby calcium promotes calcium release from intracellular stores. As we shall discuss in the following the model is inspired to the formulation \cite{Li-Hou} and set up so to make contact, in the mean field, with the celebrated model proposed by Goldbeter and collaborators \cite{goldbeter2}. We will in particular consider two species, that we shall respectively denote $Z$ and $Y$. $Z$ stands for the calcium ions $Ca^{2+}$ which are populating the cytosol, the liquid found inside cells. $Y$ is meant to label the $Ca^{2+}$ which are stored inside a specific compartment, insensitive to the IP3 and termed $\mathcal{Y}$. We will indicate with $s$ the number of  ions of type $Z$, i.e. dispersed in the cytoplasmic matrix. The integer $q$ quantifies instead 
 the abundance of species $Y$, the ions sequestered in the compartment.  

To progress in the model definition, we assume that the stochastic dynamics, which ultimately governs the evolution of the intracellular calcium, is an homogeneous Markov process. We will moreover label with $V$ the volume of the cell. $V$ defines in turn the characteristic size of the system. As we shall make clear in the following, the continuum deterministic limit is recovered by taking $V \rightarrow \infty$. 

A $Ca^{2+}$ ion can for instance migrate outside the cell, passing through specific channels which are hosted on the membrane walls. In term of chemical equation, one can ideally represent this event as:  
$$Z \xrightarrow[]{k }  0,$$
where the parameter $k$ stands for the reaction rate associated to the hypothesized transformation. Conversely, calcium ions can  reach the cytosol, coming from a second IP3 sensitive compartment, called $\mathcal{X}$. Following the CICR paradigm, this latter process is autocatalytic and can be represented as: 
$$Z  \xrightarrow[]{\nu_{1}\beta V/s} Z+Z.$$

Elements of type $Z$ can also come from the exterior of the cell, a process exemplified by:
$$0  \xrightarrow[]{\nu_0 } Z,$$

To complete the formulation of the model, we also assume that $Ca^{2+}$ can exit the IP3-insensitive compartment $\mathcal{Y}$ to increment the population of cytosolic calcium according to the reactions \footnote{We will not indulge further on elaborating on a possible biological interpretation of the  model. We rather insist on the fact that this is one of the possible microscopic, hence intrinsically stochastic formulation, which yields in the continuum limit to the aforementioned Golbeter model, as we will substantiate in the following. We have in particular decided to use two distinct chemical equations to model the $Ca^{2+}$ release, as it was done in \cite{Li-Hou}. Alternatively, one could have mimicked the process by requiring.  $Y  \xrightarrow[]{\nu_3 V/q} Z$. The general conclusion that we will derive holds also if the latter choice is instead made. Similar considerations also apply to the last two pairs of reactions.}   
\begin{equation}
\label{chem_1}
0  \xrightarrow[]{\nu_3 } Z \qquad Y \xrightarrow[]{\nu_3 V/ q}  0 
\end{equation}

The leaky transport form the IP3 insensitive pool $\mathcal{Y}$ to the cytosol is here modeled as \footnote{Although equations (\ref{chem_1}) and (\ref{chem_2}) share the same structure, they do refer to distinct molecular processes characterized by rather different reaction constants.}  :
\begin{equation}
\label{chem_2}
Y \xrightarrow[]{k_{f}}  0 \qquad 0 \xrightarrow[]{k_{f}q/V} Z
\end{equation}

Finally, the ions can take the inverse path from the cytosol to the container $\mathcal{Y}$:
\begin{equation}
Z \xrightarrow[]{\nu_2 V/s} 0 \qquad 0 \xrightarrow[]{\nu_2 } Y.
\end{equation}

We wish to emphasize again that to each chemical equations introduced above, we have attached a quantity, constant or function of the concentration $q/V$ and $s/V$, which quantifies the probability per unit of time for the reaction to eventually occur. The parameters $\nu_0$, $\nu_1\beta$, $\nu_3$ and $\nu_2$, $k$ ,$k_f$ are bound to a microscopic, although artificial, description of the scrutinized process but will be later on shown to correspond to the control parameters that appear in the deterministic model pioneered by Goldbeter. More precisely, $\beta$, $\nu_0$, $\nu_1$, $k_f$ e $k$ are positive constants, $\beta$ controlling the degree of stimulation. The functions $\nu_{2}$ and $\nu_{3}$ are respectively 
associated to the pumping process and to the release of calcium from the intracellular store. 
Following \cite{goldbeter3}, to take into account the cooperative nature of the two processes, as well as the positive feedback exerted on the transport by cytosolic $Ca^{2+}$, we posit: 
\begin{equation}
\label{nu}
\nu_{2}\Big(\frac{s}{V}\Big)=V_{M2}\dfrac{(s/V)^{n}}{K^{n}_{2}+(s/V)^{n}} \qquad \nu_{3}\Big(\frac{s}{V},\frac{q}{V}\Big)=V_{M3}\dfrac{(q/V)^{m}}{K^{m}_{R}+(q/V)^{m}}\dfrac{(s/V)^{p}}{K_{A}^{p}+(s/V)^{p}}
\end{equation}
where $V_{M2}$, $V_{M3}$ denote the maximum rates of $Ca^{2+}$ pumping into and release from the intracellular store. These processes are assumed to be mimicked by Hill like functions with cooperative indices respectively equal to $n$ and $m$. The integer index $p$  accounts instead for the degree of cooperation of the activation process. 
$K_A$, $K_2$ and $K_R$ are threshold constants for pumping, release and activation.  

When it comes to the stochastic model, the state of the system at time $t$ is known once the two integer quantities $(s,q)$ are being assigned. The analysed process is intrinsically stochastic and, as such, can be rigorously described in terms of a Master equation for the probability $P(s,q,t)$ of seeing the system in the state $(s,q)$ at the time of observation $t$. As a preliminary step, one needs to explicitly write down the transition rates
$T(\cdot|\cdot)$ from a given initial state (right entry) to the final state (left entry), compatible with the former, as dictated by the above chemical equations. In formulae one gets:
\begin{center}
\begin{equation}
\begin{split}
\label{rate}
&T(s-1,q|s,q)=k\dfrac{s}{V}\\
&T(s+1,q|s,q)=\nu_{0}+\nu_{1}\beta+\nu_{3}\left( \dfrac{s}{V},\dfrac{q}{V}\right)+k_{f}\dfrac{q}{V}\\
&T(s-1,q|s,q)=\nu_{2}\left( \dfrac{s}{V}\right)+k\dfrac{s}{V}\\
&T(s,q+1|s,q)=\nu_{2}\left( \dfrac{s}{V}\right) \\
&T(s,q-1|s,q)=\nu_{3}\left( \dfrac{s}{V},\dfrac{q}{V}\right)+k_{f}\dfrac{q}{V}.\\
\end{split}
\end{equation}
\end{center}

The Master equation that rules the dynamics of the stochastic process under the Markov assumption can be cast in the form:
\begin{center}
\begin{equation}
\begin{split}
\label{MasterEq}
\frac{\partial}{\partial t}P(s,q,t) =& -T(s+1,q|s,q)P(s,q,t)+T(s,q|s-1,q)P(s-1,q,t)\\
&-T(s-1 ,q|s,q)P(s,q,t)+T(s,q|s+1,q)P(s+1,q,t)\\
& -T(s,q+1|s,q)P(s,q,t)+T(s,q|s,q-1)P(s,q-1,t)\\ 
&-T(s,q-1|s,q)P(s,q,t)+T(s,q|s,q+1)P(s,q+1,t).\\
\end{split}
\end{equation}
\end{center}
Equation (\ref{MasterEq}) can be written in a slightly more compact form: 
\begin{equation}
\begin{split}
\label{compatta}
\frac{\partial}{\partial t}P(s,q,t) =& \Big[(\varepsilon^{+}_{s}-1)T(s-1,q|s,q)+(\varepsilon^{-}_{s}-1)T(s+1,q|s,q)\\
 +&(\varepsilon^{+}_{q}-1)T(s,q-1|s,q)+(\varepsilon^{-}_{q}-1)T(s,q+1|s,q)\Big]P(s,q,t),\\ 
\end{split} 
\end{equation}
where use has been made of the so called step operators $\varepsilon_{q}^{\pm}$, $\varepsilon_{s}^{\pm}$ defined as:
$$
\varepsilon_{s}^{\pm}f(s,q)\equiv f(s\pm 1,q)
$$
$$
\varepsilon_{q}^{\pm}f(s,q)\equiv f(s,q\pm 1).
$$

The master equation provides an exact description of the stochastic dynamics. It is however difficult to handle it analytically. Progress in the analysis can be made via perturbative calculations to which we will refer in the forthcoming section. Alternatively, the investigated system can be numerically simulated. By combining numerical and analytical tools, it is indeed possible to elaborate on the crucial role played by the stochastic fluctuations, stemming from the finite size and therefore intrinsic to the system. Before turning to discuss this important aspect, which constitute the core of the paper, we devote the remaining part of this section to deriving the mean field limit of the model, namely the underlying deterministic picture that can be formally recovered when operating in the thermodynamic limit $V \rightarrow \infty$. 

To this end, we look after the average concentrations $\left\langle s\right\rangle $ and  $\left\langle q\right\rangle$ respectively defined as:
\begin{equation*}
\left\langle s \right\rangle =\sum_{s,q}sP(s,q,t)\qquad \left\langle q\right\rangle =\sum_{s,q}qP(s,q,t).
\end{equation*}
Here the sums run over all positive integer pairs ($s$,$q$). A closed system of equations for the above quantities can be derived starting from the Master equation (\ref{MasterEq}) and following a standard procedure that we will here detail with reference to $\left\langle s \right\rangle $. 
Let us start by multiplying both members of (\ref{MasterEq}) times $s$ and sum over all possible states $(s,q)$.\\
The left hand side of the Master equation takes the form:
$$
\sum_{s,q}s\frac{d P(s,q,t)}{d t}=\dfrac{d}{d\tau}\sum_{s,q}\dfrac{s}{V}P(s,q,t)=\dfrac{d<s>}{d\tau},
$$
where we have introduced the rescaled time $\tau=t/V$.\\
Consider now the first two terms on the right hand side. By implementing in the second term the change of variable $s-1\rightarrow s$ one gets:
\begin{align*}
\sum_{s,q}s&\Big(-T(s+1,q|s,q)P(s,q,t)+T(s,q|s-1,q)P(s-1,q,t)\Big)=\\
&=\sum_{s,q}\Big(-sT(s+1,q|s,q)+(s+1)T(s+1,q|s,q)\Big)P(s,q,t)=\\
&=\left\langle T(s+1,q|s,q)\right\rangle .
\end{align*}
Similar considerations apply to the other two terms that appear in the right hand side of equation (\ref{MasterEq}):
\begin{align*}
\sum_{s,q}s&\Big(-T(s-1,q|s,q)P(s,q,t)+T(s,q|s+1,q)P(s+1,q,t)\Big)=\\
&=\sum_{s,q}\Big(-sT(s-1,q|s,q)+(s-1)T(s-1,q|s,q)\Big)P(s,q,t)=\\
&=-\left\langle T(s-1,q|s,q)\right\rangle, 
\end{align*}
where the second element of the sum has been transformed by operating the shift $s+1\rightarrow s$. The remaining terms in the Master equation are associated to changes in the species $q$ and yield no contribution to the equation for $\left\langle s \right\rangle $. To clarify this point, 
let us consider the third pair of terms in the right hand side of equation (\ref{MasterEq}). By replacing in the last of these terms $q-1\rightarrow q$ one gets:
\begin{align*}
\sum_{s,q}s&\Big(-T(s,q+1|s,q)P(s,q,t)+T(s,q|s,q-1)P(s,q-1,t)\Big)=\\
&=\sum_{s,q}\Big(-sT(s,q+1|s,q)+sT(s,q+1|s,q)\Big)P(s,q,t)=0. 
\end{align*}

Summing up, by collecting all terms together, the following equation for the average concentration $\left\langle s \right\rangle$ is eventually found:
$$
\dfrac{d<s>}{d\tau}=\left\langle T(s+1,q|s,q)\right\rangle-\left\langle T(s-1,q|s,q)\right\rangle.    
$$
By recalling the expression for the transition rates, as given in equations (\ref{rate}), we obtain:
$$
\dfrac{d<s>}{d\tau}=\nu_{0}+\nu_{1}\beta+\left\langle\nu_{3}\left( \dfrac{s}{V},\dfrac{q}{V}\right) \right\rangle+k_{f}\dfrac{\left\langle q\right\rangle }{V}-\left\langle\nu_{2}\left( \dfrac{s}{V}\right) \right\rangle-k\dfrac{\left\langle s\right\rangle }{V}.
$$
Consider for instance $\left\langle\nu_{3}\left( \dfrac{s}{V},\dfrac{q}{V}\right) \right\rangle$. In the limit $V\longrightarrow\infty$, it is legitimate to neglect the correlations which formally implies setting: 
$$\left\langle\nu_{3}\left( \dfrac{s}{V},\dfrac{q}{V}\right) \right\rangle\longrightarrow\nu_{3}\left( \dfrac{\left\langle s\right\rangle}{V},\dfrac{\left\langle q\right\rangle}{V}\right).$$ 
In conclusion, by introducing: 
\begin{eqnarray}
\label{media}
\phi &=& \lim_{V\to\infty}\left\langle s\right\rangle/V \\
\psi &=& \lim_{V\to\infty}\left\langle q\right\rangle/V
\end{eqnarray}
and recalling eqs. (\ref{nu}) one gets:
\begin{equation}
\label{cm1}
\dfrac{d \phi}{d \tau}=\nu_{0}+\nu_{1}\beta-\nu_{2}(\phi)-k\phi+k_{f}\psi+\nu_{3}(\phi,\psi).
\end{equation}
A formal identical calculation can be carried out for the other species to eventually obtain:
\begin{equation}
\label{cm2}
\dfrac{d\psi}{d\tau}=-k_{f}\psi-\nu_{3}(\phi,\psi)+\nu_{2}(\phi).
\end{equation}

\begin{figure}
\begin{center}
\includegraphics[scale=0.5]{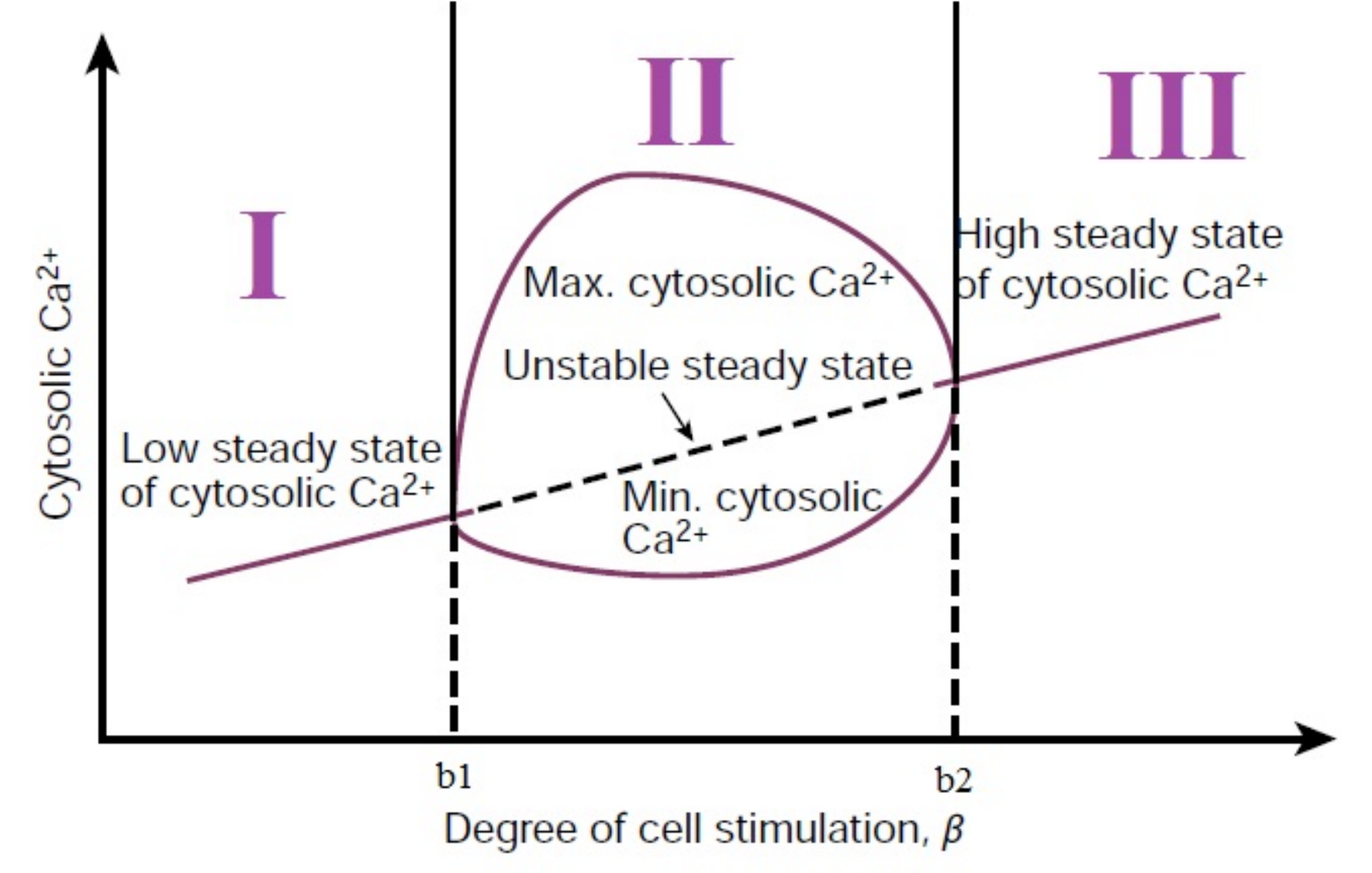}
 \caption{Schematic bifurcation diagram showing the domain and amplitude of
intracellular $Ca^{2+}$ oscillations as a function of the degree of cellular 
stimulation $\beta$, which acts as control parameter. Sustained 
oscillations develop in region II, for a range of stimulation laying between two critical
values of $\beta$,  respectively denoted $b_{1}$ and $b_{2}$.
In region I and III, the system converges to stable stationary fixed point. The solid lines in region II stand 
for lowest and highest levels of cytosolic $Ca^{2+}$ oscillations.  
 }
 \label{fig1}
\end{center} 
\end{figure}

Equations (\ref{cm1}) and (\ref{cm2}) constitute the deterministic approximation of the stochastic model. As anticipated, they match the classical model studied by Dupont and Goldbeter in \cite{goldbeter2}. Such model displays a Hopf bifurcation: by tuning the control parameter $\beta$, a transition occurs which changes the stable stationary points into an oscillating solution. Three regimes can be in particular identified, depending on the value of the degree of cell stimulation $\beta$ , and are schematically depicted in figure \ref{fig1}, adapted from \cite{goldbeterNature}. Region II, delimited by the critical values $\beta=b_{1}$ and $\beta=b_{2}$, identifies the domain where self-sustained oscillations of intracellular $Ca^{2+}$ are predicted to occur as follows a straightforward linear stability analysis applied to system (\ref{cm1})-(\ref{cm2}). In regions I and III, the concentration of $Ca^{2+}$ converges to a stationary stable state. The asymptotic concentration in
 creases linearly with the parameter $\beta$. For this reason, zones I and III are often referred to as to the regions of, respectively, low and high $Ca^{2+}$ concentration.

\begin{figure}
\begin{center}
\includegraphics[scale=0.5]{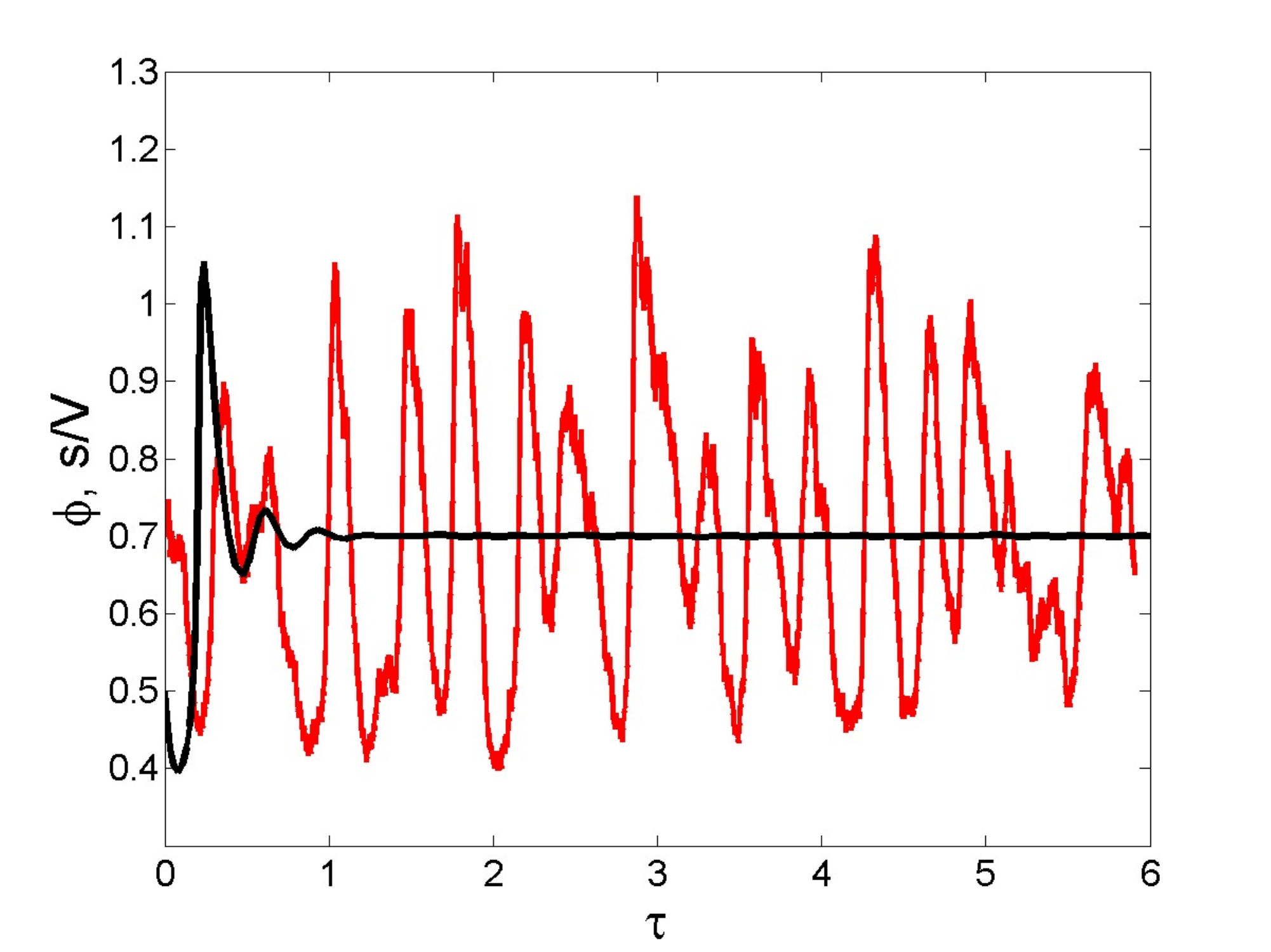}
 \caption{Time evolution of the intracellular calcium concentration $\phi$. Parameters are set so to have the system in region III, as identified in the main body of the paper and in the caption of figure 
\ref{fig1}. The black solid line which converges to an asymptotic stable fixed point refers to the integration of the deterministic system (\ref{cm1}) and (\ref{cm2}). The wiggling curve (red online) follows stochastic simulations. Persistent oscillations are found, which display a characteristic frequency.  This latter can be analytically predicted as discussed in section \ref{S3}. Parameters are $\beta=0.823$,  $K_{R}=2\mu mol/L$, $k=10s^{-1}$, $K_{A}=0.9\mu mol/L$, $n=2$, $m=2$, $p=4$, $V_{M3}=500\mu mol/(Ls)$, $V_{M2}=65\mu mol/(Ls)$, $k_{f}=1 s^{-1}$, $\nu_0=1 (\mu mol)/(Ls)$, $\nu_1=7.3 \mu mol/(Ls)$, $K_2=1 \mu mol/L $. For this choice of the parameters, the critical values of $\beta$ are respectively $b_1 = 0.291$ and $b_2 = 0.775$.
}
 \label{fig2}
\end{center} 
\end{figure}

Beyond the mean field prediction is instructive to simulate the Master equation (\ref{MasterEq}), which provides an exact description of the underlying stochastic dynamics. This task can be accomplished by resorting to the celebrated Gillespie scheme \cite{Gillespie}, a Monte Carlo based algorithm which produces realizations of the stochastic model which agree with the corresponding Master equation. In figure \ref{fig2} deterministic and stochastic simulations are confronted for a choice of the parameters that would position the system in region III. The deterministic solution (black online) approaches the asymptotic state, after an oscillatory transient that gets rapidly damped. At variance, persistent oscillations are observed in the 
stochastic model. Such sustained oscillations, also termed in the literature quasi-cycles, result from a resonant effect and originates from the amplification of the inherent finite sizes fluctuations \footnote{Notice that for the quasi cycles to develop the associated mean field system has to approach its equilibrium stationary state via damped oscillations, as it happens in region III. }. Interestingly, $Ca^{2+}$ oscillations can hence develop outside the region of the parameters for which a deterministic limit cycle is predicted to occur. This observation was already made in \cite{Li-Hou}, based on simulative evidences. 
In this paper, we shall take one step forward by characterizing the phenomenon analytically and so making contact with the 
concept of quasi-cycles as introduced above. To this end, we will compute the power spectrum of fluctuations and identify the, spontaneously selected, resonant frequency. The next section is entirely devoted to reporting about the calculations.

\subsection{The role of finite size fluctuations: characterising the stochastic oscillations.} 
\label{S3}

Consider the discrete concentration  $\dfrac{s}{V}$. Following the linear noise approximation, also called the van Kampen ansatz \cite{Vankampen}, one can express $s/V$ (resp. $q/V$) as the sum of two distinct contributions. On the one side the deterministic solution, namely  $\phi(t)$ (resp. $\psi(t)$), which denotes the concentration  in the mean field limit. The other contribution refers instead to the stochastic perturbation termed  $\xi$ (resp. $\eta$) and
is assumed to scale as $1/\sqrt{V}$. In formulae:
\begin{equation}
\label{ansazVK}
\frac{s(t)}{V}=\phi(t)+\frac{\xi}{\sqrt{V}},
\end{equation}
\begin{equation}
\label{ansazVK2}
\frac{q(t)}{V}=\psi(t)+\frac{\eta}{\sqrt{V}}.
\end{equation}
In the limit for $V \rightarrow \infty$ the stochastic contributions drop away and one is left with the deterministic concentrations $\phi$ and $\psi$. Working at finite, although large $V$, one can carry out a perturbative expansion of the governing Master equation, the quantity $1/\sqrt{V}$ acting as a small parameter. This is the van Kampen system size expansion \cite{Vankampen}, that allows one to recover the mean field equations at the leading order, and then characterize the distribution of fluctuations, at the next to leading order.  

Let us start by noting that the step operators that appear in the left hand side of eq. (\ref{compatta}), can be expanded as: 
\begin{equation*}
\varepsilon_{s}^{\pm}\simeq 1\pm\dfrac{1}{\sqrt{V}}\dfrac{\partial}{\partial \xi}+\dfrac{1}{2V}\dfrac{\partial^{2}}{\partial \xi^{2}}\qquad
\varepsilon_{q}^{\pm}\simeq 1\pm\dfrac{1}{\sqrt{V}}\dfrac{\partial}{\partial \eta}+\dfrac{1}{2V}\dfrac{\partial^{2}}{\partial \eta^{2}}.
\end{equation*}
It is then necessary to expand in series of $1/\sqrt{V}$ the transition rates, which are non-linear functions of the discrete concentrations. We shall hereafter outline the main step of the calculation with reference to the term  
 $T(s,q+1|s,q)$, and then generalize the results to the other contributions. As a first step, let us introduce the van Kampen ansatz into the formula for $T(s,q+1|s,q)$ as defined by (\ref{rate}) and (\ref{nu}):
$$
T(s,q+1|s,q)=V_{M2}\dfrac{z^{n}}{K^{n}_{2}+z^{n}}=V_{M2}\dfrac{\left( \phi+\dfrac{\xi}{\sqrt{V}}\right) ^{n}}{K^{n}_{2}+\left( \phi+\dfrac{\xi}{\sqrt{V}}\right) ^{n}}=V_{M2}\dfrac{\phi^{n}\left( 1+\dfrac{\xi}{\phi\sqrt{V}}\right) ^{n}}{K^{n}_{2}+\phi^{n}\left(1+\dfrac{\xi}{\phi\sqrt{V}}\right)^{n}}.
$$
A simple algebraic manipulation yields to: 
\begin{equation*}
T(s,q+1|s,q) \simeq V_{M2}\dfrac{\phi^{n}}{K^{n}_{2}+\phi^{n}}\left( 1+n\dfrac{\xi}{\phi\sqrt{V}}\right)\left( 1-n\dfrac{\xi}{\phi\sqrt{V}}\dfrac{\phi^{n}}{K_{2}^{n}+\phi^{n}}\right) +o\left( \dfrac{1}{\sqrt{V}}\right), 
\end{equation*}
where use has been made of the approximate relations $(1+\epsilon)^{n}=1+n\epsilon+o(\epsilon)$ and $\dfrac{1}{(1+\epsilon)}=1-\epsilon+o(\epsilon)$, for $\epsilon \ll 1$. In conclusion, neglecting again terms of order $1/V$,
one gets the final expression:
$$
T(s,q+1|s,q)\simeq V_{M2}\dfrac{\phi^n}{K^{n}_{2}+\phi^{n}}\left[ 1+\dfrac{\xi}{\sqrt{V}}\dfrac{nK^{n}_{2}}{\phi(K^{n}_{2}+\phi^{n})} \right]+o\left( \dfrac{1}{\sqrt{V}}\right) . 
$$
Following a similar strategy for the other transition rates, one eventually obtains: 
\begin{align*}
T(s+1,q|s,q)\simeq&\nu_{0}+\nu_{1}\beta+V_{M3}\dfrac{\psi^{m}}{K^{m}_{R}+\psi^{m}}\dfrac{\phi^{p}}{K^{p}_{A}+\phi^{p}}\Bigg[ 1+\dfrac{\eta}{\sqrt{V}}\dfrac{mK^{m}_{R}}{\psi(K^{m}_{R}+\psi^{m})} + \\
+&\dfrac{\xi}{\sqrt{V}}\dfrac{pK^{p}_{A}}{\phi(K^{p}_{A}+\psi^{p})} \Bigg]+k_{f}\left( \psi+\dfrac{\eta}{\sqrt{V}}\right),\\
T(s-1,q|s,q)\simeq&V_{M2}\dfrac{\phi^n}{K^{n}_{2}+\phi^{n}}\Bigg[ 1+\dfrac{\xi}{\sqrt{V}}\dfrac{nK^{n}_{2}}{\phi(K^{n}_{2}+\phi^{n})} \Bigg]+k\left( \phi+\dfrac{\xi}{\sqrt{V}}\right),\\
T(s,q+1|s,q)\simeq&V_{M2}\dfrac{\phi^n}{K^{n}_{2}+\phi^{n}}\left[ 1+\dfrac{\xi}{\sqrt{V}}\dfrac{nK^{n}_{2}}{\phi(K^{n}_{2}+\phi^{n})} \right], \\
T(s,q-1|s,q)\simeq& k_{f}\left( \psi+\dfrac{\eta}{\sqrt{V}}\right)+V_{M3}\dfrac{\psi^{m}}{K^{m}_{R}+\psi^{m}}\dfrac{\phi^{p}}{K^{p}_{A}+\phi^{p}}\Bigg[ 1+\dfrac{\eta}{\sqrt{V}}\dfrac{mK^{m}_{R}}{\psi(K^{m}_{R}+\psi^{m})}+\\
+ &\dfrac{\xi}{\sqrt{V}}\dfrac{pK^{p}_{A}}{\phi(K^{p}_{A}+\psi^{p})} \Bigg].
\end{align*}

Introduce now $\Pi(\xi,\eta,t)$, the distribution of fluctuations formally defined as:
\begin{equation}
\label{pi}
\Pi(\xi,\eta,t)=P\left( s(\phi(t),\xi),q(\psi(t),\eta),t\right), 
\end{equation}
where $ s(\phi(t),\xi)$  and $q(\psi(t),\eta)$ are respectively defined according to eqs.\ (\ref{ansazVK}) and (\ref{ansazVK2}). Taking the derivative of eq. (\ref{pi}) with respect to time yields: 
\begin{equation*}
\dfrac{\partial\Pi}{\partial t}=\dfrac{dP}{dt}=\dfrac{\partial P}{\partial t}+\dfrac{\partial P}{\partial s}V\dot{\phi}(t)+\dfrac{\partial P}{\partial q}V\dot{\psi}(t).
\end{equation*}
Hence:\\
\begin{equation}
\label{eqP}
\dfrac{\partial P}{\partial t}=\dfrac{\partial\Pi}{\partial t}-\dfrac{\partial P}{\partial s}V\dot{\phi}(t)-\dfrac{\partial P}{\partial q}V\dot{\psi}(t).
\end{equation}
On the other hand:
\begin{equation*}
\dfrac{\partial \Pi}{\partial \xi}=\dfrac{\partial P}{\partial s}\dfrac{\partial s}{\partial \xi}=\sqrt{V}\dfrac{\partial P}{\partial s}\qquad
\dfrac{\partial \Pi}{\partial \eta}=\dfrac{\partial P}{\partial q}\dfrac{\partial q}{\partial \eta}=\sqrt{V}\dfrac{\partial P}{\partial q},
\end{equation*}
which takes us to:
\begin{equation*}
\dfrac{\partial P}{\partial s}=\dfrac{1}{\sqrt{V}}\dfrac{\partial \Pi}{\partial \xi} \qquad \dfrac{\partial P}{\partial q}=\dfrac{1}{\sqrt{V}}\dfrac{\partial \Pi}{\partial \eta}.
\end{equation*}
Equation (\ref{eqP}) can be therefore cast into the form: 
\begin{equation}
\label{dpi}
\dfrac{\partial P}{\partial t}=\dfrac{\partial\Pi}{\partial t}-\dfrac{\partial \Pi}{\partial \xi}\sqrt{V}\dot{\phi}(t)-\dfrac{\partial \Pi}{\partial \eta}\sqrt{V}\dot{\psi}(t),
\end{equation}
which transform into: 
$$
\dfrac{\partial P}{\partial t}\longrightarrow\dfrac{1}{V}\dfrac{\partial\Pi}{\partial \tau}-\dfrac{1}{\sqrt{V}}\dfrac{\partial \Pi}{\partial \xi}\dot{\phi}(\tau)-\dfrac{1}{\sqrt{V}}\dfrac{\partial \Pi}{\partial \eta}\dot{\psi}(\tau).
$$
by operating the change of variable $\tau \rightarrow t/V$.
To proceed in the analysis one needs to insert into the Master equation (\ref{MasterEq}) the approximate expressions for the transition rates, as well as the above relation for  $\partial P/ \partial t$. 
The terms can be therefore re-organized  
depending on their respective order in $1/\sqrt{V}$. At the leading order, namely the terms proportional to $1/\sqrt{V}$, one eventually recovers the mean field deterministic system for the continuum densities $\phi$ and $\psi$. 
At the next to leading order, a Fokker-Planck for the distribution of the finite size fluctuations is instead obtained.

\subsection{Leading order: the mean field limit.}
The contributions relative to $1/\sqrt{V}$ result in: 
\begin{multline*}
-\dfrac{\partial \Pi}{\partial \xi}\dfrac{d\phi}{d\tau}-\dfrac{\partial \Pi}{\partial \eta}\dfrac{d\psi}{d\tau} = 
\Big(-\nu_{0}-\nu_{1}\beta-\nu_{3}(\phi,\psi)-k_{f}\psi+\nu_{2}(\phi)+k\phi\Big)\dfrac{\partial \Pi}{\partial \xi}+
\\
+\Big(k_{f}\psi+\nu_{3}(\phi,\psi)-\nu_{2}(\phi)\Big)\dfrac{\partial\Pi}{\partial\eta}.
\end{multline*}
By grouping together the terms proportional to $\dfrac{\partial \Pi}{\partial \xi}$ (resp. $\dfrac{\partial \Pi}{\partial \eta}$) and requiring their sum to return zero, one ends up with the following system of differential equations for the mean field concentrations  $\phi$ and $\psi$: 

\begin{eqnarray*}
\dfrac{d\phi}{d\tau}&=&\nu_{0}+\nu_{1}\beta+\nu_{3}(\phi,\psi)+k_{f}\psi-\nu_{2}(\phi)-k\phi\\
\dfrac{d\psi}{d\tau}&=&-k_{f}\psi-\nu_{3}(\phi,\psi)+\nu_{2}(\phi).
\end{eqnarray*}

The above equations are identical to eqs. (\ref{cm1}) and (\ref{cm2}) as derived in the preceding section. However, the van Kampen expansion enables us to take one step forward in the study of the stochastic model. A rather complete characterization of the fluctuations can be in fact gained by operating at the next to leading approximation, as we shall outline in the remaining part of this section.

\subsection{The next to leading approximation: a Fokker-Planck equation for the fluctuations.}

Consider now the terms that scale as $V^{-1}$ in the expansion of the Master equation. In formulae one has: 
\begin{align*}
&\dfrac{\partial\Pi(\xi,\eta,\tau)}{\partial \tau}
=\Bigg\{\dfrac{1}{2}\Big(k_{f}\psi+\nu_{3}(\phi,\psi)+\nu_{2}(\phi)\Big)\partial_{\eta}^{2}+\dfrac{1}{2}\Big(\nu_{0}+\nu_{1}\beta+\nu_{3}(\phi,\psi)+k_{f}\psi+\nu_{2}(\phi)+ k\phi\Big)\partial_{\xi}^{2}\Bigg\}  \Pi
\\
&+\partial_{\eta} \Bigg[\Big(k_{f}+\nu_{3}(\phi,\psi)\dfrac{mK^{m}_{R}}{\psi\Big(K^{m}_{R}+\psi^{m}\Big)}\Big)\eta+\Big(\nu_{3}(\phi,\psi)\dfrac{pK^{p}_{A}}{\phi\Big(K^{p}_{A}+\phi^{p}\Big)}-\nu_{2}(\phi)
\dfrac{nK^{n}_{2}}{\phi\Big(K^{n}_{2}+\phi^{n}\Big)}\Big)\xi\Bigg] \Pi
\\
&+\partial_{\xi} \Bigg[-\Big(k_{f}+\nu_{3}(\phi,\psi)\dfrac{mK^{m}_{R}}{\psi\Big(K^{m}_{R}+\psi^{m}\Big)}\Big)\eta
+\Big(k+\nu_{2}(\phi)\dfrac{nK^{n}_{2}}{\phi\Big(K^{n}_{2}+\phi^{n}\Big)}-\nu_{3}(\phi,\psi)\dfrac{pK^{p}_{A}}{\phi\Big(K^{p}_{A}+\phi^{p}\Big)}\Big)\xi\Bigg] \Pi.
\end{align*}
This is a linear Fokker-Planck equation, which can be put in the standard form: 
\begin{equation}
\label{FP}
\dfrac{\partial\Pi(\bold{x},\tau)}{\partial \tau}=-\sum^{2}_{i=1} \dfrac{\partial}{\partial x_{i}}A_{i}(\bold{x})\Pi(\bold{x},\tau)+\dfrac{1}{2}\sum^{2}_{i,j=1}\dfrac{\partial^{2}}{\partial x_{i}\partial x_{j}}B_{i,j}\Pi(\bold x,\tau),
\end{equation}
where we have introduced the vector $\bold x=(x_{1},x_{2})=(\xi,\eta)$. In the above eq.\ (\ref{FP}), $A_{i}(\bold x)$ 
represents the $i$-th component of the vector:

$$
A(\bold x)=M\bold x, 
$$
where $M$ is a $2 \times 2$ matrix:
\begin{equation}
\label{deriva}
M=\left( 
\begin{array}{cc}
-k-\nu_{2}(\phi)\dfrac{nK^{n}_{2}}{\phi\Big(K^{n}_{2}+{\phi}^{n}\Big)}+\nu_{3}(\phi,\psi)\dfrac{pK^{p}_{A}}{\phi\Big(K^{p}_{A}+{\phi}^{p}\Big)} &   k_{f}+\nu_{3}(\phi,\psi)\dfrac{mK^{m}_{R}}{\psi\Big(K^{m}_{R}+{\psi}^{m}\Big)} \\ 
-\nu_{3}(\phi,\psi)\dfrac{pK^{p}_{A}}{\phi\Big(K^{p}_{A}+{\phi}^{p}\Big)}+\nu_{2}(\phi)\dfrac{nK^{n}_{2}}{\phi\Big(K^{n}_{2}+{\phi}^{n}\Big)} &   -k_{f}-\nu_{3}(\phi,\psi)\dfrac{mK^{m}_{R}}{\psi\Big(K^{m}_{R}+{\psi}^{m}\Big)}
\end{array}
\right).
\end{equation}
The terms $B_{i,j}$ in equation (\ref{FP}) are the entries of the diagonal diffusion matrix $B$:
\begin{equation}
\label{diffusione}
B=\left( 
\begin{array}{cc}
\nu_{0}+\nu_{1}\beta+\nu_{3}(\phi,\psi)+\nu_{2}(\phi)+k_{f}\psi+k\phi & 0 \\ 
0 & k_{f}\psi+\nu_{3}(\phi,\psi)+\nu_{2}(\phi)
\end{array}
\right). 
\end{equation}

Notice that the coefficients of the above matrices $M$ and $B$ depend on time $\tau$, as the continuum concentration  
$\phi$ and $\psi$ do.  The Fokker-Planck equation that we have derived makes it possible to characterize the distribution of fluctuations and explains, on solid interpretative ground, the emergence of the quasi-cycles 
as reported in figure \ref{fig2}. 
To this end, by building on the general approach first derived in \cite{McKanePRL} and later on exploited in 
e.g. \cite{dipatti}, we will hereafter obtain a closed expression for the power spectrum of the stochastic fluctuations. It will be hence possible to determine a priori, and as a function of the parameters of the model, the frequency of the $Ca^{2+}$ oscillations resulting from the resonant amplification of the intrinsic noise.

\section{The power spectrum of fluctuations} 

A stochastic differential equation of the Langevin type can be associated to the Fokker-Planck equation \cite{Vankampen}. The Langevin equation describes the time evolution of the fluctuations, returning a global distribution which obeys the corresponding Fokker-Planck equation. For our case, the relevant Langevin equation, equivalent to eq. (\ref{FP}), reads:
\begin{equation}
\label{Langevin}
\dfrac{d}{d\tau}\bold{x}_{l}(\tau)=\sum^{2}_{j=1} M_{l,j}\bold{x}_{j}(\tau)+\lambda_{l}(\tau) \qquad l=1,2;
\end{equation}
where $\bold{x}_{l}$ is the $l$-th component of the vector $\bold{x}=(\xi,\eta)$ and $\lambda_{l}(\tau)$ stands for a stochastic process which satisfies the following conditions:
$$
\left\langle \lambda_{l}(\tau) \right\rangle=0 \qquad  \left\langle \lambda_{l}(\tau)\lambda_{j}(\tau') \right\rangle=B_{l,j}\delta(\tau-\tau'). 
$$
It should be noticed that the amplitude of  the noise term is controlled by the diffusion matrix $B$ and ultimately relates to the chemical parameters of the model. In other words, the noise follows the microscopic formulation of the problem and it is not imposed as an external source of disturbance.     
 
To study the emergence of regular patterns in time, the quasi-cycles, it is convenient to Fourier transform the Langevin equation (\ref{Langevin}):
\begin{equation}
\label{lan}
-\imath\omega\widehat{x}_{l}(\omega)=\sum^{2}_{j=1} M_{l,j}\widehat{x}_{j}(\omega)+\widehat{\lambda}_{l}(\omega) \qquad l=1,2;
\end{equation}
where $\widehat{\cdot}$ stands for the Fourier transform and $\omega$ represents the Fourier frequency.
The stochastic process $\widehat{\lambda}(\omega)$ verifies:
\begin{equation}
\label{lambda}
\left\langle \widehat{\lambda}_{l}(\omega) \right\rangle=0 \qquad  \left\langle  \widehat{\lambda}_{l}(\omega) \widehat{\lambda}^{*}_{j}(\omega)\right\rangle=B_{l,j},
\end{equation}
where $ \widehat{\lambda}^{*}_{j}(\omega)$ denotes the complex conjugate of   $\widehat{\lambda}_{j}(\omega)$. Eq. (\ref{lan}) yields: 
$$
\widehat{x}_{l}(\omega)=\sum^{2}_{j=1}\left( -\imath\omega I_{l,j}-M_{l,j}\right)^{-1}\widehat{\lambda}_{j}(\omega)=\sum^{2}_{j=1}\Phi^{-1}_{l,j}(\omega)\widehat{\lambda}_{j}(\omega)  \qquad l=1,2; 
$$
where $\Phi(\omega)=-\imath\omega I-M$. The power spectrum can be calculated as: 
\begin{align*}
P_{l}(\omega)=&\left\langle|\widehat{x}_{l}(\omega)|^{2}\right\rangle=\left\langle \widehat{x}_{l}(\omega)\widehat{x}^{*}_{l}(\omega)\right\rangle=\left\langle\sum^{2}_{j,r=1}\Phi^{-1}_{l,j}(\omega)\widehat{\lambda}_{j}(\omega)(\Phi_{l,r}^{*})^{-1}(\omega)\widehat{\lambda}^{*}_{r}(\omega)\right\rangle=\\
=&\sum^{2}_{j,r=1}\Phi^{-1}_{l,j}(\omega)\left\langle \widehat{\lambda}_{j}(\omega)\widehat{\lambda}^{*}_{r}(\omega)\right\rangle(\Phi^{\dagger})_{l,r}^{-1}(\omega)=\sum^{2}_{j,r=1}\Phi^{-1}_{l,j}(\omega)B_{j,r}(\Phi^{\dagger})_{l,r}^{-1}(\omega)  
\end{align*}
where use has been made of eq. (\ref{lambda}) and where we have introduced $\Phi^{\dagger}=(\Phi^{*})^{T}$.\\
In conclusion, one gets:
\begin{equation}
\label{ps}
P_{l}(\omega)=\sum^{2}_{j,r=1}\Phi^{-1}_{l,j}(\omega)B_{j,r}(\Phi^{\dagger})_{l,r}^{-1}(\omega) \qquad l=1,2. 
\end{equation}
 After some algebraic manipulation, Eq. (\ref{ps}) yields to the following explicit forms: 
\begin{eqnarray}
\label{ps_esp}
P_{Z}(\omega)=\dfrac{a_{Z}+b_{Z}\omega^{2}}{(\omega^{2}-\Omega^{2})^{2}+\Gamma^{2}\omega^{2}}\\\vspace{1cm}
P_{Y}(\omega)=\dfrac{a_{Y}+b_{Y}\omega^{2}}{(\omega^{2}-\Omega^{2})^{2}+\Gamma^{2}\omega^{2}},
\end{eqnarray}
where $\Gamma=-tr(M)$, $\Omega=\sqrt{det(M)}$  $a_{Z}=B_{1,1}M^{2}_{2,2}-2B_{1,2}M_{1,2}M_{2,2}+B_{2,2}M^{2}_{1,2}$, $b_{Z}=B_{1,1}$, $b_{Y}=B_{2,2}$ e  $a_{Y}=B_{2,2}M^{2}_{1,1}-2B_{1,2}M_{2,1}M_{1,1}+B_{1,1}M^{2}_{2,1}$.

\begin{figure}[!h]
\begin{center}
\includegraphics[scale=0.5]{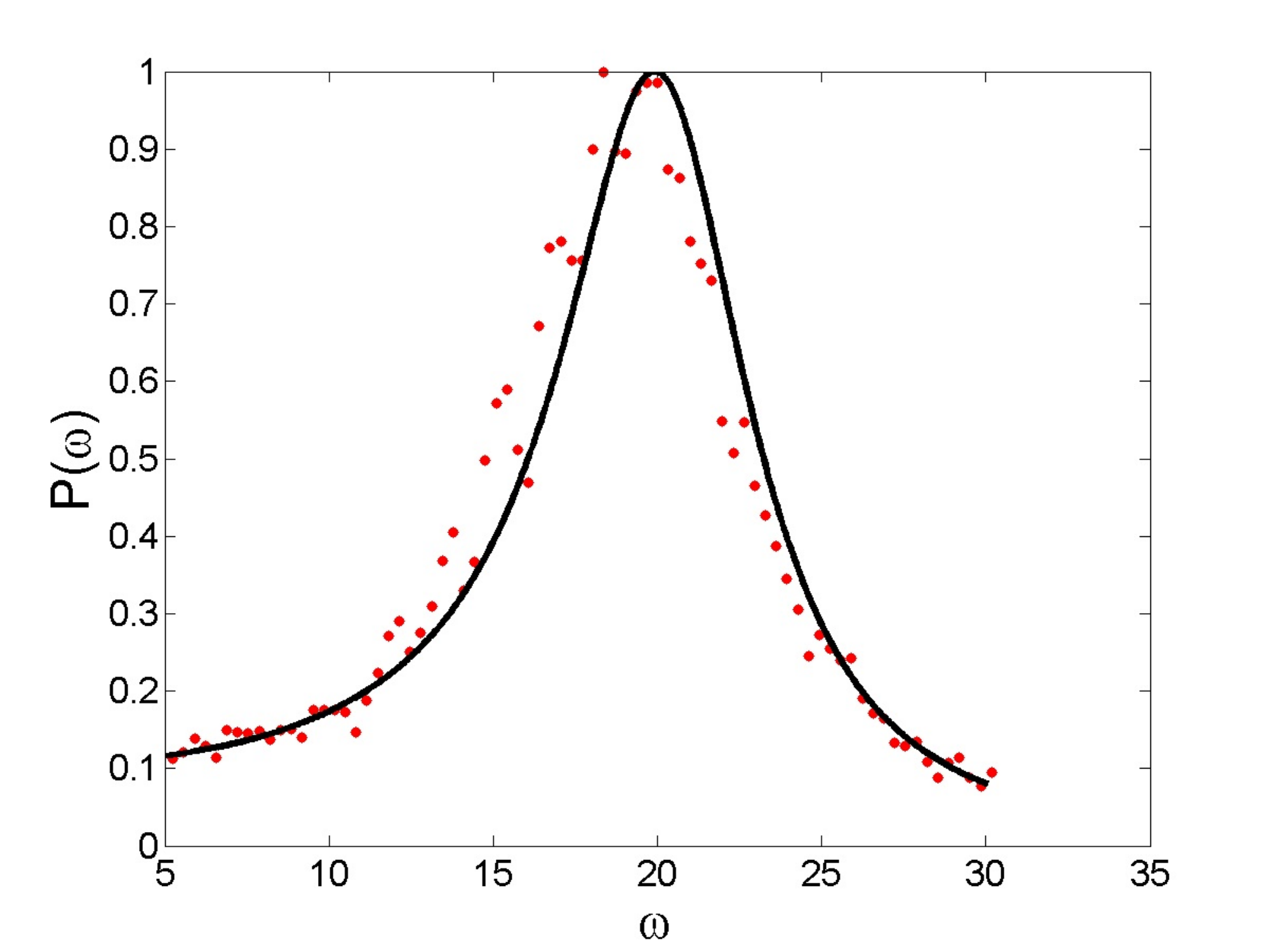}
\end{center}
\caption{\small{The power spectrum of fluctuations as a function of $\omega$ is plotted for species $Z$. The solid line stands for the theoretical prediction (\ref{ps_esp}), while the symbols refer to the stochastic simulations averaged over $200$ independent realizations. The parameters are set as in  figure \ref{fig2}. The system is hence initialized to fall in region III, as depicted in the schematic diagram of figure \ref{fig1}.
}}
\label{ps2}
\end{figure}

To test the adequacy of the theory we can plot the power spectrum of fluctuation of species $Z$ around the mean field stationary point, for a choice of the parameters that falls outside the region of deterministic oscillations and compare the prediction to the numerical profile obtained by averaging over many realizations of the stochastic dynamics. The comparison is displayed in figure \ref{ps2}. A clear peak is found in the power spectrum, thus confirming that the stochastic oscillations as seen in figure \ref{fig2} stem from finite size fluctuations. The agreement between theory and numerical simulations is excellent. Quasi cycles can therefore develop outside the region of mean field order, and the associated frequency can be adequately estimated via 
perturbative analytical means.  

An approximate estimate for the resonant frequency, where the peak of the power spectrum is positioned, can be analytically worked out. Let us assume legitimate to neglect, as a first approximation, the term proportional to $\omega^2$ in the numerator of $P_{Z}(\omega)$, see  eq. (\ref{ps_esp}). Then $P_{Z}(\omega)$ is maximum when the denominator is minimum, namely when $|det\Phi(\omega)|$ takes the smallest possible value. Denote by  $\lambda_{i}$, $i=1,2$, the eigenvalues of matrix $M$. Hence:
$$
|det\Phi(\omega)|=\prod^{2}_{j=1}(-\imath\omega-\lambda_{j})(\imath\omega-\lambda^{*}_{j}).
$$  
Since $M$ is by definition a real matrix, the eigenvalues $\lambda_{i}$ can be either real or complex conjugate. 
If they are real, then $|det\Phi(\omega)|=(\omega^{2}+\lambda_{1}^{2})(\omega^{2}+\lambda_{2}^{2})$.
If they are complex $\lambda_{1}=\lambda^{*}_{2}=\lambda$. If one posits $\lambda=\lambda_{R}+\imath\lambda_{I}$, then: 
\begin{equation}
\label{omega}
|det\Phi(\omega)|=|\omega^{2}+(\lambda_{R}^{2}-\lambda^{2}_{I})+2\imath\lambda_{R}\lambda_{I}|^{2}.
\end{equation}

This latter case is of interest to us, as quasi-cycles can develop only if the corresponding mean field dynamics approaches the asymptotic stationary state, via damped oscillations. The condition for the minimum of (\ref{omega}) readily translates into 
the final expression for the resonant frequency $\omega_{max}$: 

\begin{equation}
\label{omega_max}
\omega_{max}=\sqrt{\lambda^{2}_{I}-\lambda^{2}_{R}}.
\end{equation}

\begin{figure}[!h]
\begin{center}
\includegraphics[scale=0.5]{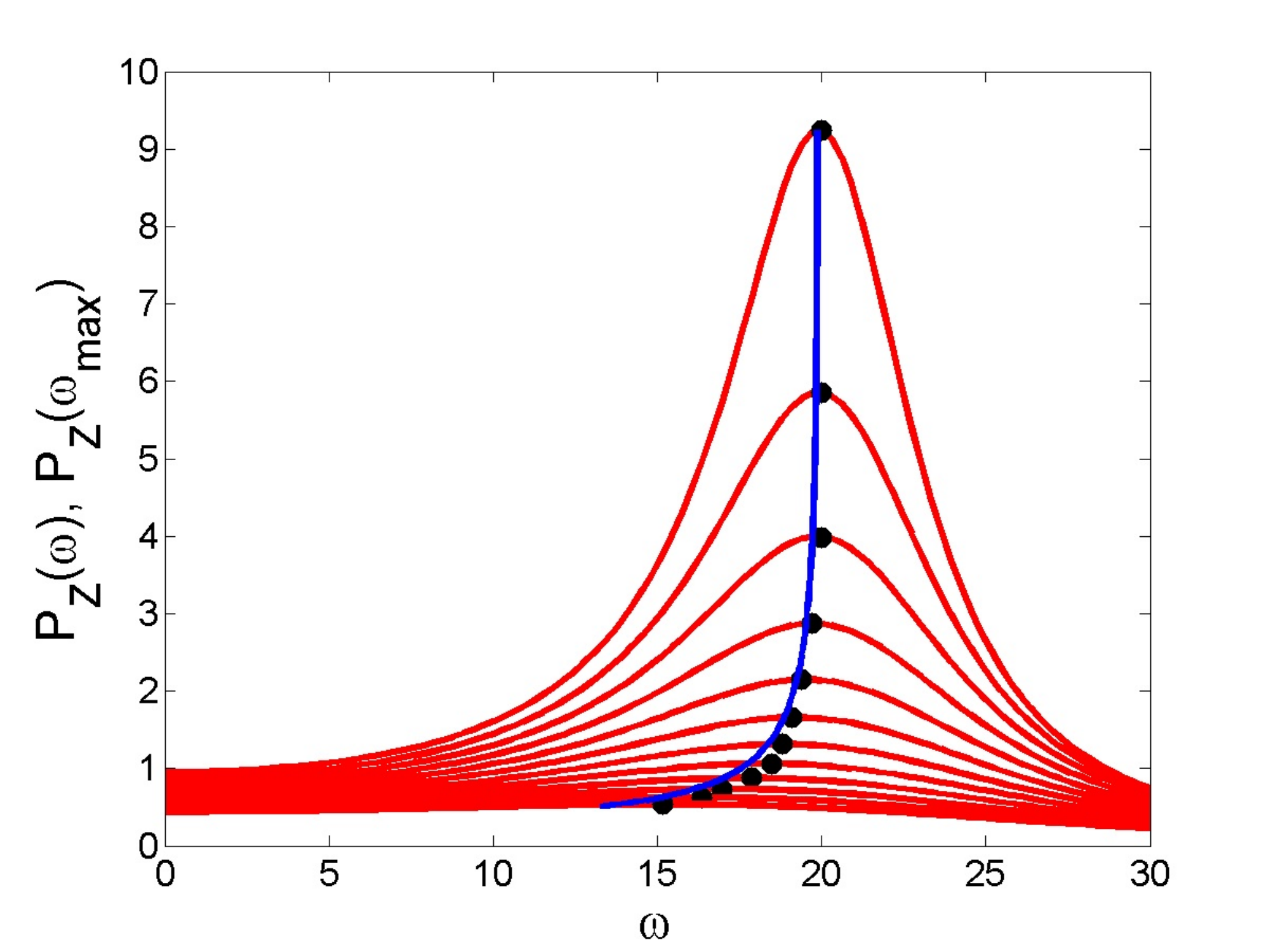}
\end{center}
\caption{\small{The power spectrum of fluctuations $P_Z$ as a function of $\omega$, for different choices of $\beta$. is plotted for species $Z$. The symbols identify the location of the maxima, while the solid line is obtained by plotting 
$P_Z(\omega_{max})$, as a function of $\beta$. The quantity $\omega_{max}$ follows the approximate eq. (\ref{omega_max}). 
}}
\label{fig4}
\end{figure}

In figure \ref{fig4} the theoretical power spectrum of fluctuations is plotted for different values of $\beta$. The symbols identify the position of the peaks while the solid line refers to the approximate formula 
$P_Z(\omega_{max})$, where $\omega_{max}$ is given by (\ref{omega_max}). This comparison points to the correctness of formula 
(\ref{omega_max}), which therefore encodes all the necessary information to estimate the resonant frequency as a function of the parameters of the model. 

\section{Conclusion} 
\label{S4}

Calcium oscillations are crucial for the functioning of the cellular machinery and for this reasons have been widely investigated, both experimentally and theoretically. Dynamical models have been proposed, to different levels of sophistication, which make it possible to reproduce in silico the processes that  underly the emergence of sustained 
$Ca^{2+}$ oscillations. Most of the models so far discussed in the literature are of deterministic inspiration and, as such, omit the inherent stochastic perturbations, which stem from individual based effects. This latter are particularly important at low concentrations, a regime which is certainly of interest when it comes to modeling intracellular calcium oscillations. 

Recently, Li, Hou and Xin \cite{Li-Hou} have put forward a microscopic version of the celebrated Goldbeter model \cite{goldbeter1, goldbeter2}, a paradigmatic scheme for calcium dynamics, which assumes two species in mutual interaction. Working in such a  generalized stochastic setting, it was shown numerically \cite{Li-Hou} that the intrinsic noise can drive stochastic oscillations, termed in \cite{McKanePRL} quasi cycles,  also outside the region where a deterministic limit cycle is predicted to occur. 

Starting from this observation, and to make contact with the theoretical literature devoted to the phenomenon of quasi-cycles,  we have here carried out an analytical study of a stochastic $Ca^{2+}$ model \cite{Li-Hou}. The analysis is carried out under the linear noise approximation and allows us to obtain a close prediction for the power spectrum of stochastic fluctuations. This latter displays an isolated peak, whose reference frequency appears to be controlled by the chemical parameters of the model, as e.g. the degree of external stimulation $\beta$. The validity of the theory is confirmed by direct numerical simulations of the examined stochastic model.

In conclusion, by building on recent advances on the study of noise induced oscillations in stochastic population dynamics models, we have here cast on solid mathematical ground the numerical observations of \cite{Li-Hou}, so confirming that intrinsic noise can play an important, although often neglected, role in the onset of intracellular calcium oscillations.

\end{document}